\documentclass[aps,prl,twocolumn,floatfix,superscriptaddress,showpacs]{revtex4}
\usepackage[dvips]{color}
\usepackage{graphicx}
\usepackage{amsmath,amssymb,bm}
\usepackage[hypertex,breaklinks=true,colorlinks=false]{hyperref}

\newcommand{\Tr}{\mathrm{Tr}}
\newcommand{\CC}{\mathrm{CC}}

\newcommand{\bra}[1]{\langle #1|}
\newcommand{\ket}[1]{|#1 \rangle}

\newcommand{\Ab}{{\bar{A}}}

\begin{document}
\title{
Mutual Information and Boson Radius in $c=1$ Critical Systems in One Dimension
}
\author{Shunsuke Furukawa}
\affiliation{Condensed Matter Theory Laboratory, RIKEN, Wako, Saitama 351-0198, Japan}
\author{Vincent Pasquier}
\affiliation{Institut de Physique Th\'eorique, CEA Saclay, 91191 Gif-sur-Yvette Cedex, France}
\author{Jun'ichi Shiraishi}
\affiliation{Graduate School of Mathematical Science, University of Tokyo, Komaba, Meguro-ku, Tokyo 153-8914, Japan}

\date{\today}
\begin{abstract}
We study the generic scaling properties of the mutual information between two disjoint intervals, 
in a class of one-dimensional quantum critical systems 
described by the $c=1$ bosonic field theory. 
A numerical analysis of a spin-chain model reveals 
that the mutual information is scale-invariant 
and depends directly on the boson 
radius. 
We interpret the results in terms of correlation functions of branch-point twist fields.
The present study provides a new way to determine the boson 
radius, 
and furthermore demonstrates the power of the mutual information 
to extract more refined information 
of conformal field theory 
than the central charge. 
\end{abstract}

\pacs{03.67.Mn, 05.70.Jk, 75.10.Pq}


\maketitle

Conformal field theory (CFT) provides a powerful framework 
to study one-dimensional (1D) quantum many-body systems. 
One can recast the low-energy degrees of freedom of both the bosonic and fermionic gases 
into a simple bosonic field theory, 
known as the Tomonaga-Luttinger liquid (TLL) 
\cite{Haldane81,Giamarchi04}. 
Thanks to the recent technological advances, 
a precise correspondence between the TLL predictions and various (quasi-)1D systems, such as 
carbon nanotubes \cite{Ishii03}, 
antiferromagnetic chains \cite{Klanjsek08}, 
and cold Bose gases in 1D traps \cite{Kinoshita04} is currently being investigated.





Given a microscopic model, 
an important and often nontrivial issue is how
to obtain the effective field theory 
controlling its long-distance behavior. 
The notion of quantum entanglement, or more specifically, the entanglement entropy, 
has been extensively applied
as a new way to address this basic matter.
From a quantum ground state $|\Psi\rangle$, 
one constructs the reduced density matrix $\rho_A:=\Tr_\Ab ~\ket{\Psi}\bra{\Psi}$ on a subsystem $A$ 
by tracing out the exterior $\Ab$. 
The entanglement entropy is defined as $S_A:=-\Tr ~\rho_A \log \rho_A$.
In 1D quantum critical systems, 
the entanglement entropy for an interval $A=[x_1,x_2]$ embedded in a chain 
exhibits a universal scaling 
\cite{Holzhey94,Vidal03,Jin04,Calabrese04,Laflorencie06,Ryu06}:
\begin{equation}\label{eq:S_A}
 S_A = 
 (c/3) \log (x_2-x_1) + s_1, 
\end{equation}
where $c$ is the central charge of the CFT and $s_1$ is a non-universal constant related to the ultra-violet (UV) cutoff.
This scaling allows to determine the universal number $c$ 
as a representative of the ground state structure, 
without having to worry about the precise correspondence between the microscopic model 
and the field theory.

As it is well known, the central charge is not the only important number specifying a CFT. 
In the bosonic field theory with $c=1$, 
the boson compactification radius $R$ 
(or equivalently, the TLL parameter $K=1/(4\pi R^2)$) 
is a dimensionless parameter 
which changes continuously in a phase 
and controls the power-law behavior of various physical quantities. 
It is natural to ask  {\em  how  to identify the boson radius as a generic structure of the ground state}. 
In this Letter, we demonstrate that the entanglement entropy can  achieve this task 
if we consider two disjoint intervals, $A=[x_1,x_2]$ and $B=[x_3,x_4]$.
We analyze the scaling of the mutual information defined as 
\begin{equation}\label{eq:MI}
 I_{A:B} := S_A+S_B-S_{A\cup B}.
\end{equation}
This measures the amount of information shared by two subsystems \cite{Adami97,Vedral97}. 
A numerical analysis of a spin-chain model 
reveals a robust relation between $I_{A:B}$ and $R$, 
irrespective of microscopic details. 
We compare the result with the general prediction of Calabrese and Cardy (CC) \cite{Calabrese04}, 
and find a relevant correction to their result. 


Roughly speaking, the mutual information~\eqref{eq:MI} may be regarded 
as a {\it region-region} correlator. 
It is known that $I_{A:B}$ is non-negative, and becomes zero 
iff $\rho_{A\cup B}=\rho_A \otimes \rho_B$, 
i.e., in a situation of no correlation \cite{Cheong08}.
A motivation to consider $I_{A:B}$ comes from 
that microscopic details at short-range scales, 
which are often obstacles when analyzing {\it point-point} correlators, 
can be smoothed out over regions. 
As we enlarge the region sizes, 
we expect that $I_{A:B}$ detects essential features of the correlations emerging in the coarse-grained limit.
When there is a long-range order in local operators, 
we have $I_{A:B}\ne 0$ for finite local regions $A$ and $B$, 
even in the limit of large separation \cite{DegEnt}.
In a critical system with power-law decaying correlations, 
$I_{A:B}$ goes to zero if $A$ and $B$ are  far apart in comparison with their lengths, $r_A$ and $r_B$. 
However, if  $r_A$ and $r_B$ are of the order of the separation, 
$I_{A:B}$ can remain finite, which is the situation we examine here.

\begin{figure}[t]
\begin{center}
\includegraphics[width=8cm]{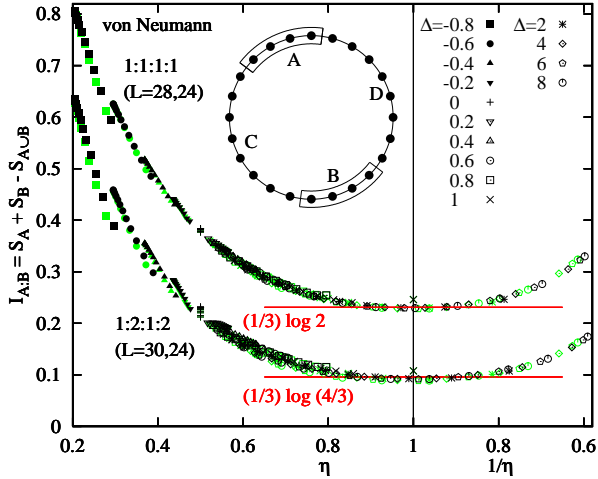}
\end{center}
\caption{(color online)
The mutual information for fixed divisions 
$r_A$:$r_C$:$r_B$:$r_D$=1:1:1:1 and 1:2:1:2, 
versus $\eta=2\pi R^2$.
We set the magnetization  at $M=\frac{k}{L}$
with $k=0,1,\dots,\frac{L}2-3$ for $-1<\Delta\le 1$
and with $k=1,\dots,\frac{L}2-3$ for $1<\Delta$, 
so that the system is inside the critical phase.
Black and green points correspond to
the larger ($L=28,30$) and smaller ($L=24$) systems, respectively.
Horizontal red lines 
indicate the Calabrese-Cardy result \eqref{eq:MI_CC}.
}
\label{fig:MI_R}
\end{figure}

First, suppose we treat the mutual information \eqref{eq:MI} 
following the prediction of Calabrese and Cardy \cite{Calabrese04}, 
which will turn out in our analysis to correspond to the $SU(2)$-symmetric case.  
For an infinite chain, 
the entanglement entropy on double intervals $A\cup B$
was predicted to be \cite{Calabrese04}
\begin{equation}\label{eq:S_AB_CC}
 S_{A\cup B} = \frac{c}3 \log \left( \frac{x_{21}x_{32}x_{43}x_{41}}{x_{31}x_{42}} \right) + 2s_1,
\end{equation}
with $x_{ij}=x_i-x_j$. 
Here the constant term $2s_1$ is determined
so that $S_{A\cup B}\to S_A + S_B$ in the limit $x_{21},x_{43} \ll x_{31},x_{42}$. 
For a finite chain of length $L$, 
one replaces $x_{ij}$ by the cord distance 
$
 \frac{L}{\pi} \sin \frac{\pi x_{ij}}{L}
$
in Eqs.~\eqref{eq:S_A} and \eqref{eq:S_AB_CC} \cite{Calabrese04,Ryu06}. 
We now consider a division $(r_A,r_C,r_B,r_D)$ of a finite chain 
as depicted in the inset of Fig.~\ref{fig:MI_R}. 
Then the CC formula for the mutual information reads
\begin{equation}\label{eq:MI_CC}
 I_{A:B}^\CC
 = \frac{c}{3} \log \left[
    \frac{\sin\frac{\pi(r_A+r_C)}{L} \sin\frac{\pi(r_B+r_C)}{L}}{\sin\frac{\pi r_C}{L} \sin\frac{\pi r_D}{L}}
                    \right].
\end{equation}
Note that the UV-divergent constant $s_1$ has been cancelled out
in the mutual information, and the resultant 
\eqref{eq:MI_CC} is invariant under global scale 
transformations. 
Similar ideas of eliminating the
UV-divergence have been suggested by Casini and Huerta \cite{Casini04} and
have also been exploited in the context of topological entropy
\cite{Kitaev06} in higher dimensions. 
Henceforth, lengths of (sub)systems are measured in units of the lattice spacing. 

Now we turn to numerical analyses of the mutual information in a spin chain,
based on Lanczos diagonalization of finite systems up to $L=30$.
We consider a spin-$\frac12$ XXZ chain in a magnetic field,
\begin{equation}
  H:=\sum_{j=1}^{L} (S^x_j S^x_{j+1} + S^y_j S^y_{j+1} + \Delta S^z_j S^z_{j+1}) - h\sum_{j=1}^{L} S^z_j.
\end{equation}
Since the magnetization per site, $M:=\frac1L \sum_j S^z_j$, is a conserved quantity,
we can label the ground-states using $M$.
A $c=1$ critical phase extends over a wide region in $\Delta>-1$ \cite{Giamarchi04}.
The boson radius $R$ 
controls the nature of critical correlations.
Indeed, the exponents for the leading algebraic decay of spin correlations,
$\langle S^x_j S^x_{j^\prime} \rangle$ and $\langle S^z_j S^z_{j^\prime} \rangle-M^2$,
are given by $\eta:= 2\pi R^2$ and $\min (1/\eta,2)$, respectively. 
Henceforth we frequently use $\eta :=2\pi R^2$ instead of $R$. 
For $-1<\Delta\le 1$ and $h=0$, we have 
$
 \eta = 1-(1/\pi)\arccos \Delta.
$
For $\Delta>1$, the system is in a gapped N\'eel phase at $h=0$
and enters the critical phase at a critical field with $\eta=2$. 
For general $h\ne 0$, $\eta$ can be determined by
numerically solving the integral equations obtained from the Bethe ansatz 
\cite{Bogoliubov86,Qin97,Cabra98}. 
When increasing $h$, $\eta$ monotonically increases ($-1<\Delta<0$) or
decreases ($0<\Delta$) to $\frac12$ at the saturation. 
Summaries of the value of $\eta$ 
in the $M$-$\Delta$ and $h$-$\Delta$ phase diagrams can be found 
in e.g. Refs.~\onlinecite{Giamarchi04} and \onlinecite{Cabra98}.


We first evaluate $I_{A:B}$ for fixed divisions
$(r_A,r_C,r_B,r_D)=\frac{L}4 (1,1,1,1)$ and $\frac{L}6 (1,2,1,2)$.
Figure~\ref{fig:MI_R} shows a plot of $I_{A:B}$ against $\eta=2\pi R^2$ 
for various $(M,\Delta)$ in the critical phase.
Remarkably, the data points almost form a single curve for each type of division.
The collapse of a two-dimensional $M$-$\Delta$ plane onto these two curves
strongly indicates a direct dependence of $I_{A:B}$ on $R$.
Agreement with the CC formula \eqref{eq:MI_CC} can be observed
only around $\eta=1$ ($SU(2)$-symmetric case).
One can also observe that $I_{A:B}$ is symmetric under $\eta\to 1/\eta$,
which might reflect the duality in the effective theory.


\begin{figure}[t]
\begin{center}
\includegraphics[width=7.5cm]{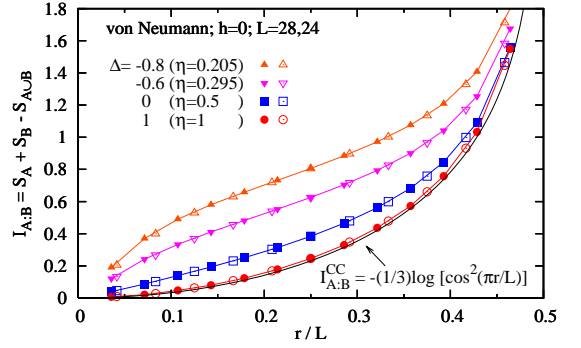}
\end{center}
\caption{(color online)
Mutual information $I_{A:B}$ as a function of $\frac{r}{L}$
for divisions $(r_A,r_C,r_B,r_D)=(r,\frac{L}2-r,r,\frac{L}2-r)$. 
We set $h=0$, and symbols with different shapes correspond to different $\Delta=-0.8,-0.6,0,1$. 
Filled and empty symbols correspond to $L=28$ and $24$, respectively.
}
\label{fig:MI_n}
\end{figure}

\begin{figure}[t]
\begin{center}
\includegraphics[width=8.7cm]{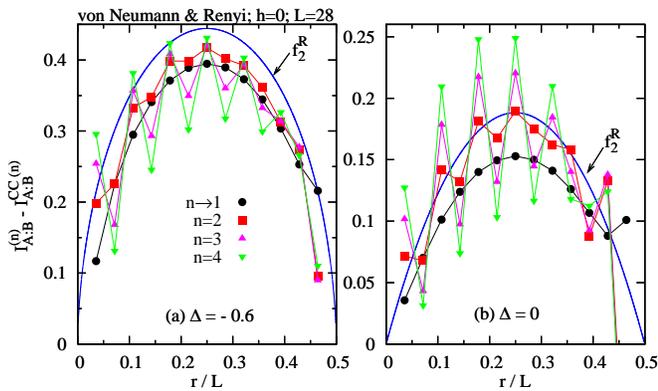}
\end{center}
\caption{(color online) 
The deviation of the ``R\'enyi'' mutual information $I_{A:B}^{(n)}$ 
from the CC result $I_{A:B}^{\CC(n)}$ 
for divisions $(r,\frac{L}2-r,r,\frac{L}2-r)$. 
Different symbols correspond to $n=1,2,3,4$. 
}
\label{fig:MIR_n}
\end{figure}

In Fig.~\ref{fig:MI_n},
we plot $I_{A:B}$ as a function of $\frac{r}{L}$
for divisions $(r_A,r_C,r_B,r_D)=(r,\frac{L}2-r,r,\frac{L}2-r)$,
in comparison with CC formula~\eqref{eq:MI_CC}.
For each $\Delta$, the results from $L=28$ and $24$ obey a single curve, 
indicating the scale invariance of $I_{A:B}$. 
The curve for $\Delta=1$ agrees well with the CC formula~\eqref{eq:MI_CC}. 
In other cases, the curves run above the CC formula. 
We can confirm that $I_{A:B}$ approaches zero in the limit $\frac{r}{L}\to 0$, 
as expected for systems without long-range order.
If we subtract the CC formula (see black circles in Fig. \ref{fig:MIR_n}),
we find that the curves are symmetric under $\frac{r}{L} \to \frac12- \frac{r}{L}$
and have maxima at $\frac{r}{L}=\frac14$.


\newcommand{\Zamo}{\mathrm{Z}}

As an extension of the von Neumann entropy, 
we also consider the R\'enyi entropy (or alpha
entropy) defined as
\begin{equation}
 R_A^{(n)} := \frac{-1}{n-1} \log (\mathrm{Tr}~\rho_A^n).
\end{equation}
The von Neumann entropy $S_A$ can be reached in the limit
$n\to 1$. Following Calabrese and Cardy \cite{Calabrese04}, 
one can derive the following expression 
(originally found in Ref.~\cite{Jin04})
for a single interval $A=[x_1,x_2]$ in an infinite chain: 
\begin{equation}
 R_A^{(n)} = \frac{1+n}{6n} c~\log x_{21} + s_n,
\end{equation}
where $s_n$ is again a UV-divergent constant.
Likewise, {\em within CC argument}, 
the translation from von Neumann to R\'enyi can be done via replacements
$\frac{c}{3} \to \frac{1+n}{6n} c$, $s_1\to s_n$.
We define the ``R\'enyi'' mutual information as
$
 I_{A:B}^{(n)} := R_A^{(n)} + R_B^{(n)} - R_{A\cup B} ^{(n)}.
$

In Fig.~\ref{fig:MIR_n}, we plot the deviation of the ``R\'enyi'' mutual information $I_{A:B}^{(n)}$
from the CC prediction
$
 I_{A:B}^{\CC(n)} = - \frac{1+n}{6n} c ~\log\left[ \cos^2 \frac{\pi r}L \right].
$
In contrast to the von Neumann case $n\to 1$, we
observe some oscillating dependence on $\frac{r}{L}$ for $n>1$.
Similar oscillations have
also been reported for the single-interval entropy in Ref.~\onlinecite{Nienhuis08}. 
In Fig.~\ref{fig:MIR_n}, the oscillations in the $n=3$ and $4$ cases occur 
around the relatively smooth curves in the von Neumann case. 
From this, it is expected that 
$I_{A:B}^{(n)}-I_{A:B}^{\CC(n)}$ consists of 
{\em a smooth component}, which depends little on $n$, 
and {\em an oscillating component}, which shrinks in the limit $n\to 1$. 
The former should be controlled by the continuum description.

\begin{figure}[t]
\begin{center}
\includegraphics[width=7.5cm]{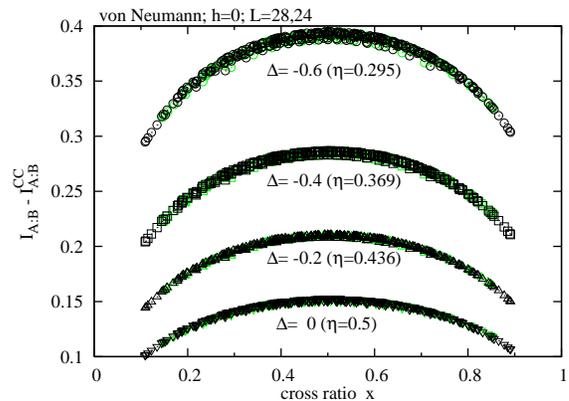}
\end{center}
\caption{(color online)
$I_{A:B}-I_{A:B}^{\CC}$ versus the cross ratio $x$ given in \eqref{eq:cr_L}.
All the divisions $(r_A,r_C,r_B,r_D)$
with $3\le r_A \le r_B$ and $3\le r_C \le r_D$ are examined.
For $L=28~(24)$, there are totally $305~(152)$ possibilities of such divisions.
Black and Green symbols correspond to $L=28$ and $24$, respectively.
}
\label{fig:MI_cr}
\end{figure}

\newcommand{\Deltab}{{\bar{\Delta}}}
\newcommand{\xb}{{\bar{x}}}
\newcommand{\Tcal}{{\mathcal T}}
\newcommand{\Tcalt}{\tilde{\Tcal}}

Let us now discuss the origins of 
the deviation from the Calabrese-Cardy prediction and 
the dependence on the boson radius. 
We follow the formulation based on
branch-point twist fields proposed by Cardy {\it et al.}~\cite{Cardy07} 
First, we represent the moment $\Tr~ \rho_A^n$ as the partition
function on a $n$-sheeted Riemann surface ${\cal R}_n$
\cite{Calabrese04}. Then, we relate it to a correlation function
of twist fields $\Tcal$ and $\Tcalt$ with conformal dimensions
$\Delta_n=\Deltab_n= \frac{c}{24} \left(n-\frac1n \right)$ \cite{Cardy07}. For
double intervals $A\cup B = [x_1,x_2]\cup [x_3,x_4]$ in
an infinite chain, we can write it down as
\begin{align}\label{eq:twist4}
 \Tr~ \rho_{A\cup B}^n
  \propto \langle \Tcal (x_1) \Tcalt (x_2) \Tcal (x_3) \Tcalt (x_4) \rangle.
\end{align}
The $SL(2,\mathbb{C})$ covariance property requires this four-point function to have the following form:
\begin{align}\label{eq:corr4}
 \left( \frac{x_{31}x_{42}}{x_{21}x_{32}x_{43}x_{41}} \right)^{2\Delta_n}
   \left( \frac{\xb_{31}\xb_{42}}{\xb_{21}\xb_{32}\xb_{43}\xb_{41}} \right)^{2\Deltab_n}
   F_n(x,\xb; \eta),
\end{align}
with $x_i=\xb_i$. Here, $F_n(x,\xb; \eta)$ is
a function of the cross ratios 
$x:=\frac{ x_{21}x_{43} }{x_{31}x_{42} }$ and 
$\xb:=\frac{ \xb_{21}\xb_{43} }{\xb_{31}\xb_{42} }$, 
normalized as 
$\lim_{x\to 0} F_n(x,x;\eta)=1$, 
and should be determined by $\eta=2\pi R^2$
as suggested  by Fig.~\ref{fig:MI_R}. 
The power function part
$(\dots)^{2\Delta_n} (\dots)^{2\Deltab_n}$ in Eq.~\eqref{eq:corr4}
corresponds to the CC prediction \cite{Calabrese04}, and
the function $F_n$ gives an additional contribution
$\frac{-1}{n-1} \log F_n(x,x; \eta) =: -f_n(x;\eta)$ to 
the R\'enyi entropy $R^{(n)}_{A\cup B}$. The mutual information
detects this new part:
\begin{align} 
 &I_{A:B}^{(n)}-I_{A:B}^{\CC(n)}=f_n(x;\eta),\label{eq:MIR_fn}\\
 &I_{A:B}-I_{A:B}^{\CC}=\lim_{n\to 1} f_n(x;\eta) =: f(x;\eta).\label{eq:MI_fn}
\end{align}
The function $f_n(x;\eta)$ should satisfy 
(i) $f_n(x;\eta)\to 0~(x\to 0)$, 
(ii) the crossing invariance $f_n(x;\eta)=f_n(1-x;\eta)$
required from $R^{(n)}_{A\cup B}=R^{(n)}_{C\cup D}$ for a finite chain
(see Eq. \eqref{eq:cr_L} below),
and (iii) $f_n(x;\eta)=f_n(x;1/\eta)$ and $f_n(x;1)=0$ suggested by Fig.~\ref{fig:MI_R}.


As a check of this result, we plot $I_{A:B}-I_{A:B}^{\CC}$ as a
function of the cross ratio $x$ in Fig.~\ref{fig:MI_cr}. For a
finite chain, the cross ratio is given by 
\begin{equation}\label{eq:cr_L}
 x= \frac{ \sin\frac{\pi r_A      }{L} \sin\frac{\pi r_B      }{L} }
         { \sin\frac{\pi (r_A+r_C)}{L} \sin\frac{\pi (r_C+r_B)}{L} }
\end{equation}
We can confirm that for a given $\Delta$, and for various
divisions $(r_A,r_C,r_B,r_D)$, the additional contribution to the
CC result can be fit by a single curve 
with good accuracy, strongly supporting Eq. \eqref{eq:MI_fn}.


For $n=2$, two twist fields, $\Tcal$ and $\Tcalt$, are identical, 
and have conformal dimensions $\Delta_2=\Deltab_2=1/16$. 
The correlation function \eqref{eq:twist4} of four twist fields with these dimensions 
(Ramond fields) 
has been derived previously \cite{Zamolodchikov86,Dixon87}.
The crossing-invariant solution 
gives
\begin{equation}\label{eq:corr4_Ash}
 f_2^\mathrm{R} (x;\eta ) = \log \frac{\theta_3 (\eta \tau) \theta_3 (\eta^{-1} \tau)}{ [\theta_3 (\tau)]^{2}},
\end{equation}
where $\tau$ is pure-imaginary, and is related to $x$ via
$
 x= [\theta_2(\tau)/\theta_3(\tau)]^4
$.
Here $\theta_2$ and $\theta_3$ are Jacobi theta functions:
\begin{equation}
 \theta_2(\tau) := \sum_{m\in\mathbb{Z}} e^{i\pi\tau (m+1/2)^2},~~
 \theta_3(\tau) := \sum_{m\in\mathbb{Z}} e^{i\pi\tau m^2}.
\end{equation}
At a special point $\eta=1/2$, Eq.~\eqref{eq:corr4_Ash} can be simplified as 
$f_2^\mathrm{R} (x;1/2) = \log \sqrt{ (1+x^{1/2})(1+(1-x)^{1/2})/2 }$.
One can check that 
$f_2^\mathrm{R}(x;\eta)$ satisfies 
all the aforementioned properties (i)-(iii). 
When two intervals of small lengths $x_{21}=x_{43}=:r$ 
are separated far apart by a distance $x_{31}=x_{42}=: d (\gg r)$, 
Eq.~\eqref{eq:corr4_Ash} reduces to 
$f_2^\mathrm{R}
 \approx 
2 \left(\frac{r}{4d}\right)^{2\min (1/\eta,\eta)}$.
This scales as the dominant spin correlation {\it squared}.  

\begin{figure}[t]
\begin{center}
\includegraphics[width=7.5cm]{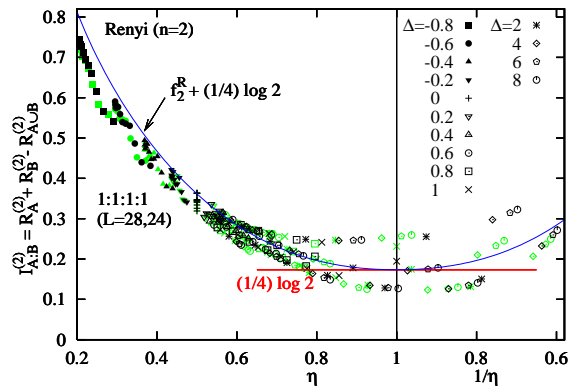}
\end{center}
\caption{(color online) 
$n=2$ ``R\'enyi'' mutual information for the 1:1:1:1 division versus $\eta=2\pi R^2$.
The same symbols as in Fig.~\ref{fig:MI_R} are used.
}
\label{fig:MIR2_R}
\end{figure}

In Fig.~\ref{fig:MIR_n}, the formula $f^\mathrm{R}_2(x;\eta)$ is drawn as smooth blue lines.
The formula agrees relatively well with the data of $I^{(2)}_{A:B}-I^{\CC(2)}_{A:B}$ for $\Delta=0$,
and runs slightly above the data for $\Delta=-0.6$.
In Fig. \ref{fig:MIR2_R}, we plot $I^{(2)}_{A:B}$ for the 1:1:1:1 division,
in comparison with $f^\mathrm{R}_2+I^{\CC(2)}_{A:B}$.
For $\eta\gtrsim 0.5$, $I^{(2)}_{A:B}$ contains strong oscillations,
and the formula goes inside these oscillations.
For $\eta\lesssim 0.5$, oscillations are small,
but the formula goes slightly above the data.
This small disagreement might be due
to finite-size effects,
to a subtle difference between lattice systems and continuum descriptions,
or to some missing factor in Eq.~\eqref{eq:corr4_Ash}.

To summarize, we have shown that the scaling of the mutual information $I_{A:B}$ 
is controlled directly by the boson radius $R$. 
This result can be used as a new method to determine $R$ from the ground state, 
complementary to the standard spectroscopic method \cite{Giamarchi04}
based on the Drude weight and the compressibility.
To obtain an analytical expression of 
$f_n(x;\eta)$ for general $n$ and especially its $n\to1$ limit $f(x;\eta)$
remains a challenging and intriguing open problem.
In general, 
we expect that every CFT has its characteristic function $f(x)$ in the mutual information. 
This can be used as a fingerprint for distinguishing different CFTs, 
as originally suggested in Ref.~\onlinecite{Casini04}.


The authors are grateful to
J.~Cardy, A.~Furusaki, 
D.~Ivanov, H.~Katsura, G. Misguich, B. Nienhuis, M.~Oshikawa, S.~Ryu, and Masahiro~Sato
for stimulating discussions.
The collaboration of the authors was initiated in the Workshop of 
``Topological Aspects of Solid State Physics'' 
at ISSP, 
Univ. of Tokyo.

{\it Note added}.
After the preprint of this paper was posted on arXiv, 
Calabrese and Cardy have added a note to their paper \cite{Calabrese04}. 
Another analysis of the double-interval entropy 
has been done in parallel by Caraglio and Gliozzi \cite{Caraglio08}.


\newcommand{\etal}{{\it et al.}}
\newcommand{\PRL}[3]{Phys. Rev. Lett. {\bf #1}, \href{http://link.aps.org/abstract/PRL/v#1/e#2}{#2} (#3)}
\newcommand{\PRLp}[3]{Phys. Rev. Lett. {\bf #1}, \href{http://link.aps.org/abstract/PRL/v#1/p#2}{#2} (#3)}
\newcommand{\PRA}[3]{Phys. Rev. A {\bf #1}, \href{http://link.aps.org/abstract/PRA/v#1/e#2}{#2} (#3)}
\newcommand{\PRAp}[3]{Phys. Rev. A {\bf #1}, \href{http://link.aps.org/abstract/PRA/v#1/p#2}{#2} (#3)}
\newcommand{\PRB}[3]{Phys. Rev. B {\bf #1}, \href{http://link.aps.org/abstract/PRB/v#1/e#2}{#2} (#3)}
\newcommand{\PRBp}[3]{Phys. Rev. B {\bf #1}, \href{http://link.aps.org/abstract/PRB/v#1/p#2}{#2} (#3)}
\newcommand{\PRBR}[3]{Phys. Rev. B {\bf #1}, \href{http://link.aps.org/abstract/PRB/v#1/e#2}{#2} (R) (#3)}
\newcommand{\arXiv}[1]{arXiv:\href{http://arxiv.org/abs/#1}{#1}}
\newcommand{\condmat}[1]{cond-mat/\href{http://arxiv.org/abs/cond-mat/#1}{#1}}
\newcommand{\JPSJ}[3]{J. Phys. Soc. Jpn. {\bf #1}, \href{http://jpsj.ipap.jp/link?JPSJ/#1/#2/}{#2} (#3)}
\newcommand{\PTPS}[3]{Prog. Theor. Phys. Suppl. {\bf #1}, \href{http://ptp.ipap.jp/link?PTPS/#1/#2/}{#2} (#3)}
\newcommand{\hreflink}[1]{\href{#1}{#1}}

\end{document}